\documentstyle[12pt]{article}

\let\slo=\o

\def\a{\alpha}
\def\b{\beta}

\def\d{\delta}
\def\f{\phi}                    %       \varphi
\def\g{\gamma}

\def\k{\kappa}
\def\l{\lambda}
\def\m{\mu}
\def\n{\nu}
\def\o{\omega}
\def\th{\theta}                  %       \vartheta
\def\r{\rho}                    %       \varrho
\def\D{\Delta}

\def\G{\Gamma}

\def\O{\Omega}

\def\cd{{\cal D}}
\def\ct{{\cal T}}
\def\cp{{\cal P}}
\def\cn{{\cal N}}

\def\rhs{\mbox{r.h.s.} }

\def\ie{\mbox{i.e.} }
\def\eg{\mbox{e.g.} }

\def\APH#1{Ann. Phys. {\bf #1}}
\def\CMP#1{Comm. Math. Phys. {\bf #1}}

\def\NPB#1{Nucl. Phys. {\bf B#1}}

\def\PLB#1{Phys. Lett. {\bf B#1}}
\def\PRD#1{Phys. Rev. {\bf D#1}}
\def\PR#1{Phys. Rev. {\bf #1}}
\def\PRL#1{Phys. Rev. Lett. {\bf #1}}

\def\MPL#1{Mod. Phys. Lett. {\bf A #1}}
\def\GR#1{Gen. Rel. and Grav. {\bf #1}}

\def\beq{\begin{equation}}
\def\eeq{\end{equation}}
\def\bqry{\begin{eqnarray}}
\def\eqry{\end{eqnarray}}

\def\det{{\rm det\,}}
\def\tr{{\rm tr\,}}

\def\Tilde#1{\widetilde{#1}}

\def\svev#1{\left\langle #1\right\rangle}
\def\vev#1{\Big\langle #1 \Big\rangle}
\def\sc{\,\raisebox{-.6ex}{$\stackrel{\textstyle\sin}{\cos}$}}

\renewcommand{\thefootnote}{\fnsymbol{footnote}}

\begin{document}
\noindent October 1994 \hfill WIS--94/44--PH
\par
\begin{center}
\vspace{15mm}
{\large\bf
     Quantum Gravity via Random Triangulations of R$^{
\mbox{\scriptsize (}\mbox{\scriptsize\bf 4}\mbox{\scriptsize )}}$ \\[2mm]
           and Gravitons as Goldstone Bosons of SL(4)/O(4)
}\\[10mm]
Yigal Shamir\footnote{
\noindent Present address:
School of Physics and Astronomy,
Tel-Aviv University, Ramat Aviv 69978, ISRAEL

\noindent email: ftshamir@wicc.weizmann.ac.il }\\[5mm]
{\it Department of Physics\\
Weizmann Institute of Science, Rehovot 76100, ISRAEL}\\[15mm]
{ABSTRACT}\\[2mm]
  \end{center}
\begin{quotation}
\begin{quotation}

  A model of random triangulations of a domain in $R^{(4)}$ is presented. The
global symmetries of the model include SL(4) transformations and translations.
If a stable microscopic scale exists for some range of parameters, the model
should be in a translation invariant phase where SL(4) is spontaneously broken
to O(4). In that phase, SL(4) Ward identities imply that the correlation
length in the spin two channel of a symmetric tensor field is infinite.
Consequently, it may be possible to identify the continuum limit of four
dimensional Quantum Gravity with points inside that phase.

\end{quotation}
\end{quotation}

\setcounter{footnote}{0}
\renewcommand{\thefootnote}{\arabic{footnote}}

\newpage

\noindent {\large\bf 1.~~Introduction and conclusions}
\vspace{3ex}

  Understanding four-dimensional Quantum Gravity ranks among the most important
open problems in current theoretical physics. In the dynamical triangulation
(DT) approach, one postulates that the metric degrees of freedom in
$d$-dimensions can be represented by a discrete sum over abstract
$d$-triangulations (for a recent review see ref.~\cite{d}). The DT approach is
particularly successful in two dimensions~\cite{dt}.  Non-trivial scaling
relations predicted by continuum techniques~\cite{2D} have been confirmed in
the DT model both analytically, by invoking its relation to matrix models, and
by extensive numerical simulations.

  The situation is different in the DT approach to four dimensional Quantum
Gravity~\cite{am,aj}. Numerical simulations based on a discretized version
of the Einstein-Hilbert action indicate the possibility of a
phase transition between a ``branched polymer'' phase and a ``crumpled''
phase~\cite{am,aj,ns}. But, in the absence of analytical support for the
numerical results, the latter should be taken with  care. In particular, it is
unclear whether or not the observed susceptibility peak signals a true phase
transition, with critical couplings that remain finite as the infinite volume
limit is taken~\cite{ckr,aj2,bs}.

  General Relativity is a theory with a local spacetime symmetry group.
It is well-known that general coordinate invariance is badly broken by any
discretization of spacetime. The solution to this problem offered by the DT
approach, is to pick a single representative for each microscopic metric in the
form of an abstract triangulation. This solution may indeed be satisfactory
in two dimensions. The reason, we believe, is related to the fact that two
dimensional gravity is a trivial theory on the
classical level. The dynamics of the quantum theory is dominated by strongly
fluctuating metrics, and its physical states bare little resemblance to smooth
two dimensional  manifolds (or to smooth manifolds of any dimensionality, for
that matter).

  On the other hand it is an experimental fact that, in four dimensions,
gravitational effects can be described accurately on a wide range of time and
distance scales by a {\it classical} theory, that is to say, by  General
Relativity. One inevitably faces the difficult question of how to recover
smooth macroscopical metrics out of the partition function of four dimensional
DT~\cite{am,bs}. Since coordinate systems play a central role in the
description of manifolds, an important problem is how to recover general
covariance of the equations of motion on macroscopic scales. Considerable work
has been devoted to these issues, but so far little progress was
made.

  What distinguishes four dimensional gravity from gravity in any lower
dimension, is the existence of a massless spin two particle associated with
the gravitational interaction -- the graviton. One way to proceed is to ask
what is the minimal subgroup of general coordinate transformations that is
still potent enough to imply the existence of the graviton, and
to try to preserve only that smaller symmetry in the discretized model.

  In a continuum framework, an answer to this question was given by
Nakanishi and Ojima~\cite{nnio}. They observed that, instead of relating the
masslessness of the graviton to general coordinate invariance, one can
attribute it to the spontaneous breakdown of a large but {\it global} spacetime
symmetry down to O(4).

  The continuum theory of ref.~\cite{nnio} is of course a gauge fixed version
of General Relativity, and the global symmetry of the gauge fixed lagrangian is
GL(4). The order parameter for GL(4) symmetry breaking is the vacuum
expectation value of the metric tensor $\svev{g^{\m\n}}\propto\d^{\m\n}$.
On a formal level, one can invoke the Goldstone theorem\footnote{
  Earlier attempts to relate the masslessness of gauge particles, namely,
the photon or the graviton, to the Goldstone theorem were made in
refs.~\cite{bj,ggb}.
}
to argue that there
should exist massless excitations associated with the generators in the coset
space GL(4)/O(4). We comment that the broken generators associated with
physical spin two gravitons actually belong to SL(4)/O(4).
The extra invariance under global scale transformations is therefore
unnecessary for this reasoning.

  General Relativity, or one of its extensions, are the only perturbatively
consistent continuum theories that contain an interacting, massless spin two
particle~\cite{gr,mr}. Consequently, as long as we remain within the context of
continuum field theory, considerations of the kind described above can only
lead us to various gauge fixed versions of the (ill-defined) continuum path
integral of General Relativity. But these considerations can
nevertheless be helpful when we try to construct a discretized model
that will reproduce General Relativity at sufficiently low energies.

  In this paper we present a model of four dimensional Quantum Gravity whose
global symmetries include SL(4) transformations and translations. The model is
based on the ensemble of geometric triangulations of $R^{(4)}$, instead of the
commonly used ensemble of abstract four-triangulations with a fixed topology.
The finite volume partition function is defined by summing over random
triangulations of a bounded domain $\cd \subset R^{(4)}$. We assume that $\cd$
is a closed, convex polyhedral domain.

  Quantum Gravity requires some discretization of spacetime. This statement
sounds so obvious, that one often forgets that it involves two distinct
aspects of the theory. A non-perturbative construction
requires the partition function's measure to be  mathematically well-defined.
The functional measure of the {\it continuum} path integral is ill-defined
mathematically, and so some discretization of the partition function's measure
is mandatory in any definition of Quantum Gravity.

  A different question is whether spacetime itself is continuous or discrete at
the smallest scale. In our model, every triangulation  has a finite number of
vertices, and the partition function's measure is mathematically
well-defined. But every vertex is identified with a point in
$R^{(4)}$ and, in this sense, a {\it continuous four-dimensional spacetime}
exists at the most fundamental level. This distinguishes the present model from
both the DT approach~\cite{am,aj} and the quantum Regge calculus
approach~\cite{qrc}. The results of this paper are due to the new measure,
which depends explicitly on the coordinates of $R^{(4)}$. At this stage we
have less understanding of the role of action, but it is unlikely that the
commonly used Regge action will be the appropriate one.

  Apart from the choice of boundary conditions, the definition of the partition
function involves the {\it vector space} structure of $R^{(4)}$, but not
its metric structure. As a result, the linear symmetry of the model is SL(4).
This can be compared with (continuum or lattice) flat space field theories,
where the linear symmetry is O(4) or a discrete subgroup of it. Now, while the
global symmetry is enhanced to SL(4), the manifest symmetry of observables
cannot exceed O(4). Hence, SL(4) has to break spontaneously.  The usual metric
structure of $R^{(4)}$ arises only after spontaneous symmetry breaking, and
this fact is what allows {\it metric fluctuations} to have a dynamical role in
the continuum limit.

  The main subject of this paper is the derivation of SL(4) Ward identities.
An obvious condition for obtaining a continuum behaviour, is that the length
of links will remain finite as the infinite volume limit is taken. The precise
condition is expressed in terms of suitable bounds on a probability
distribution that characterizes the dominant link length.
If this condition is satisfied, SL(4) is spontaneously broken to O(4),
and the dominant link length can be used to define a stable, dynamical
microscopic scale.  We believe that this is also a sufficient condition for
recovery of translation invariance in the infinite volume limit\footnote{
  A variant where translation and rotation invariance is further broken to a
discrete subgroup could also lead to a consistent continuum limit, see the
last section.
}.

 We will define below a positive definite symmetric tensor field $g^{\m\n}(x)$
whose value depends on the four-simplex containing the point $x$.  As suggested
by our notation, we propose to identify $g^{\m\n}(x)$ with the inverse metric
tensor of General Relativity. The vacuum expectation value $\svev{g^{\m\n}}$
will be the order parameter for SL(4) symmetry breaking. In the SL(4)/O(4)
broken phase, the SL(4) Ward identities imply that {\it the correlation length
in the spin two channel of} $g^{\m\n}(x)$ {\it is infinite}.  Thus, stability
of the microscopic scale defined by the dominant link length is a necessary
{\it and} sufficient condition for obtaining a non-trivial continuum behaviour.

  Conventionally, a divergent correlation length is interpreted as signaling
the presence of a massless particle. The validity of this
interpretation depends on one's ability to carry out the analytic continuation
to Minkowski space in a consistent way. Maison and Reeh~\cite{mr} proved long
ago that Goldstone bosons with non-zero spin cannot exist in a local, Lorentz
covariant theory with a positive norm Hilbert space. Therefore, when we
analytically continue back to Minkowski space, we should expect to obtain in
the
long distance limit a {\it gauge fixed} version of General Relativity.

  The implication of the Maison-Reeh theorem is that we are facing the familiar
choice between a manifestly Lorentz covariant formulation and a manifestly
unitary one.  The model defined in this paper is manifestly covariant, but it
lacks reflection positivity. In this formulation, the model cannot contain the
physical graviton and be manifestly unitary at the same time. Hence, its
validity depends ultimately on one's ability to construct a consistent set of
physical observables.  At the moment, the term ``Goldstone boson'' as used
in this paper should be understood in a loose sense, namely, as implying only
that the Fourier transform of the appropriate euclidean correlators has a
zero momentum singularity.

\vspace{5ex}
\noindent {\large\bf 2.~~The partition function}
\vspace{3ex}

   Each triangulation $\ct$ of a domain $\cd\subset R^{(4)}$ defines both
an adjacency matrix (and, hence, an abstract four-triangulation with a
boundary), and a division of $\cd$ into simplexes. Let us denote the
coordinates of the vertices of the triangulation by $y^\m_i\in\cd$,
$i=1,\ldots,N_0$. For $0\le n\le 4$, the {\it geometrical} $n$-simplex
$\D(y_{i_0},\ldots,y_{i_n}) \subset \cd$ associated with the {\it abstract}
$n$-simplex $(i_0,\ldots,i_n)$ is defined to be the set of all points
$x^\m\in\cd$ such that $x^\m=\sum_{j=0}^n \l_j y^\m_{i_j}$, where
$\sum_{j=0}^n \l_j = 1$ and $\l_j\ge 0$. We demand that
the union of all geometrical four-simplexes in $\ct$ coincide with $\cd$,
and that if $\D_1$ and $\D_2$ are geometrical simplexes that belong
to $\ct$, then the same should be true for their intersection (if it is
non-empty).

  The definition of a geometric simplex in terms of the coordinates of its
vertices makes use of the linear vector space structure of $R^{(4)}$.
It does not depend on the $R^{(4)}$ norm. A similar statement applies to
the volume of a geometric four-simplex $V=|\det W^\m_{~j}|$, where
\beq
  W^\m_{~j}(y_0,\ldots,y_4) = y^\m_0-y^\m_j \,, \quad\quad
  \m,j=1,\ldots,4\,.
\label{W}
\eeq
We will allow the partition function's measure to depends only on
$\det W$. As a result, the global symmetries of our model will
include SL(4) transformations and translations.

  The existence of an SL(4) symmetry  will play a crucial role below.
The choice of a domain in $R^{(4)}$ as the embedding space is dictated
by the need to have a well-defined partition function, where the only SL(4)
breaking effect is due to the boundary conditions. Other symmetric spaces such
as a four-sphere or a four-torus are not vector spaces, and a
knowledge of the adjacency matrix and the positions of the vertices is not
sufficient to determine a unique division of the space.  In the case of a
four-sphere, the curvature induces an explicit SL(4) breaking effect locally.
In the case of a four-torus, although locally it is indistinguishable from
$R^{(4)}$, the possibility of an arbitrary winding number makes it unclear how
to write down a well-defined partition function.  We comment in passing that,
in continuum gravity, one does not know how to define energy in the absence of
spatial boundaries.  (The electrostatic analogy is that, by Gauss law, only an
overall neutral charge distribution can fit into a cube with periodic boundary
conditions). Consequently, one should expect difficulties with the recovery of
stable single particle states, if one uses a four-sphere or a four-torus
topology.

  The partition function is defined as follows. The domain $\cd\subset R^{(4)}$
will be taken to be a hypercube $-L/2\le x^\m \le L/2$ which we denote $\cd_0$.
A fixed three-triangulation is chosen on the faces of $\cd_0$. We will define
below an (inverse) metric tensor $g^{\m\n}(x)$.  We assume that the
boundary triangulation has a discrete rotation symmetry, which ensures that the
expectation value of $g^{\m\n}(0)$ is proportional to the identity matrix.

  The partition function is
\beq
  Z = \sum_\ct e^{-S(\ct)}\, \Tilde{Z}(\ct) \,,
\label{pf}
\eeq
\beq
  \Tilde{Z}(\ct) = \prod_{i=1}^{N_0} \int_{-L/2}^{L/2} d^4y_i\,
                   \O(y_1,\ldots,y_{N_0}) \,.
\label{pftr}
\eeq

  The sum in eq.~(\ref{pf}) runs over all abstract four-triangulations that
have a realization in $\cd_0$ with the prescribed boundary triangulation.
It includes a summation over different identifications of the faces of the
abstract triangulations with the faces of $\cd_0$, and implicit in
its definition is an appropriate symmetry factor needed to avoid double
counting.

  $S(\ct)$ is an action that depends only on the adjacency matrix of
the abstract triangulation. The results of this paper
depend mainly on the choice of the {\it measure} $\O(y_1,\ldots)$, and so
the explicit form of the action will be left unspecified until the last
section. Here we only comment that one can generalize $S(\ct)$ to include
also matter fields.

  $\Tilde{Z}(\ct)$ represents an integration over all realizations of the
abstract triangulation $\ct$ with a given identification of the boundaries.
$N_0$ is the number of internal vertices of $\ct$. The term realization is used
to stress that we do not consider arbitrary embeddings of the vertices of $\ct$
in $\cd_0$. Allowed embeddings must preserve the abstract
triangulation structure. Namely, if $(i_0,\ldots,i_k)$ is
the intersection of two abstract simplexes
$(i'_0,\ldots,i'_n)$ and  $(i''_0,\ldots,i''_m)$,
then the intersection of the corresponding geometrical
simplexes $\D(i'_0,\ldots,i'_n)$ and  $\D(i''_0,\ldots,i''_m)$
should be the geometric simplex $\D(i_0,\ldots,i_k)$.

  In ref.~\cite{adf}, models of two dimensional DT embedded in some $R^{(n)}$
were considered, and it was found that area actions give rise to an
instability against the formation of infinitely long spikes.  As should be
clear from the discussion of ref.~\cite{adf}, this instability does not occur
if the dimensionality of the triangulation is equal to that of the target
space, and one does not allow for arbitrary embeddings, but only for
valid realizations of the abstract triangulation.

  The restriction to valid realizations only, is enforced analytically as
follows. Consider one realization of $\ct$. Notice that this defines an
orientation on $\ct$. For every abstract four-simplex in $\ct$,
we now fix an ordering $(i_0,\ldots,i_4)$ such  that
$\det W(y_{i_0},\ldots,y_{i_4})>0$.
The measure is defined as a product over all
four-simplexes with the prescribed ordering
\beq
  \O(y_1,\ldots,y_{N_0}) =
     \prod_{(i_0,\ldots,i_4)}
     F \left( \det W(y_{i_0},\ldots,y_{i_4})/\bar{a}^4 \right)\,,
\label{msr}
\eeq
where\footnote{
  It should be possible to choose a measure such that the product
$\prod d^4y_i\, \O(y_1,\ldots)$ will be GL(4) invariant.
However, such a scale invariant measure will have a
singular behaviour as the volume of any four-simplex tends to zero, and it
is likely to yield undesirable instabilities.
}
\beq
  F(s) = \left\{
         \begin{array}{ll}
         s\,, & s\ge 0\,, \\
         0\,, & s\le 0\,.
         \end{array} \right.
\label{F}
\eeq

   The vanishing of $F(s)$ for negative $s$, guarantees
that two geometrical four-simplexes sharing a common tetrahedron will always
lie on opposite sides of the hyper-surface defined by their common tetrahedron.
Together with the abstract triangulation conditions fulfilled by $\ct$, this
guarantees that we have a legal realization. $\bar{a}$ is a reference
microscopic scale defined below. We will usually set
$\bar{a}=1$ and omit the $\bar{a}$-dependence.
The explicit form of $F(s)$ eq.~(\ref{F}) favours
four-simplexes with a large volume. However, as we discuss below, because of
entropy considerations the actual volume distribution of four-simplexes should
be exponentially damped.

  Consider now the symmetric tensor $g^{\m\n}(x)$ defined as follows. We first
define for each geometric four-simplex
\beq
   g^{\m\n}(\D) = \sum_{<ij>} (y^\m_i-y^\m_j) (y^\n_i-y^\n_j) \,.
\label{gmn}
\eeq
The sum is over the links of the four-simplex $\D$.
Notice that $g^{\m\n}(\D)$ is a strictly positive tensor for
$\O(y_1,\ldots)\ne 0$. We now let $g^{\m\n}(x)=g^{\m\n}(\D)$ if $x$ belongs to
the interior of $\D$. This defines $g^{\m\n}(x)$ except on a subspace of
codimension one, which can be neglected under the integration in
eq.~(\ref{pftr}).

  A heuristic motivation to identify $g^{\m\n}(x)$ as defined above with the
inverse metric tensor, comes from the following observation~\cite{y}.
Let us add to $S(\ct)$  a scalar kinetic term, defined as
a sum over all links of $(\f_i-\f_j)^2$. Suppose now that $\f_i=\f(y_i)$
for some function $\f(x)$. If the approximation
$\f_i-\f_j\approx (y^\m_i-y^\m_j) \partial_\m \f$
is a valid one, then the sum of $(\f_i-\f_j)^2$ over the links of a
four-simplex takes the form of the standard continuum lagrangian for a scalar
field in a gravitational background, where the role of the inverse metric
tensor is played by $g^{\m\n}(\D)$.

  The cubic symmetries of the boundary conditions imply
$\langle g^{\m\n}(0) \rangle \propto \d^{\m\n}$ in a finite volume.
If translation invariance is recovered in the infinite volume limit,
the same will be true for
$\langle g^{\m\n}(x) \rangle = \langle g^{\m\n}(0) \rangle$.
This would imply that  the unique ground state is flat space.
As we mentioned in the introduction, we believe that
stability of the microscopic scale defined by the dominant link length, is
a necessary and sufficient condition for recovery of
translation and rotation invariance in the long distance limit.

  $g^{\m\n}(0)$ will be our order parameter for SL(4) symmetry breaking.
An SL(4) transformation acts on the coordinates of a vertex as
$y^\m_i \to (y'_i)^\m = A^\m_{~\n}\, y^\n_i$. Using eq.~(\ref{gmn}),
$g^{\m\n}(0)$ is transformed into
$(g')^{\m\n}(0) = A^\m_{~\l}\, A^\n_{~\r}\, g^{\l\r}(0)$.
In an infinitesimal form,
$\d g(0) = B\, g(0) + g(0)\, B^T$,
where $A= \exp(\a B)$. Therefore,
the expectation value $\langle g^{\m\n}(0) \rangle$ breaks SL(4) down to
O(4). The unbroken generators correspond to antisymmetric matrices, whereas
the broken generators correspond to symmetric traceless matrices.

  Before we turn to the derivation of the Ward identities, we wish to make one
more comment on the dynamics of the model. For given values of
parameters, the partition function should be dominated by
triangulations with some typical number of vertices $\bar{N}_0$. Suppose now
that we interchange the order of summation and integration in eqs.~(\ref{pf})
and~(\ref{pftr}). We then start with a given set of $\bar{N}_0$ points, and we
have to draw links connecting different pairs until the conditions of a
geometric triangulation are ultimately fulfilled.  A necessary condition for
declaring that five given points are the vertices of a geometric four-simplex,
is that the simplex will contain no other vertices of the triangulation. In
the limit of large $\bar{N}_0$, the probability that this condition is
satisfied will be proportional to $\exp(-V/\bar{a}^4)$ where
$\bar{a}^4=L^4/\bar{N}_0$.

  The probability distribution for the volume of four-simplexes $\cp(V)$ is
actually a dynamical quantity. It depends not only on the entropy factor, but
also on the explicit form of $F(s)$, as well as on the action $S(\ct)$.  We
expect its asymptotic behaviour to be $\cp(V)\sim V\exp(-V/\bar{a}^4)$. This
form should be valid at both small and large $V$, whereas at intermediate
values, the role of the action $S(\ct)$ may lead to a more complicated
behaviour.

  The volume probability distribution, taken by itself, does not place any
restrictions on the probability distribution for the length of links. The
reason is that the volume of a four-simplex is SL(4)-invariant, whereas the
length of a link is not.  Here the boundary conditions play a crucial role.
For given value of $\bar{a}$, we will
choose the boundary triangulation such that boundary tetrahedra have a regular
shape and their three-volume is $O(\bar{a}^3)$. As a result, the preferred link
length near the boundary should be $O(\bar{a})$.  Moreover, that tendency
should prevail throughout the entire volume, as long as the average number of
four-simplexes remains finite. The reason is that SL(4) is a {\it global}
symmetry, and so the free energy should increase if we change the local average
of an SL(4)-sensitive quantity such as the shape of four-simplexes.

  Whether the stabilizing effect of the boundary conditions survives in the
infinite volume limit, is the crucial dynamical question.  The SL(4)
Ward identities, to which we now turn, will clarify what are the dynamical
conditions needed for obtaining  a continuum behaviour in the infinite
volume limit.

\newpage
%\vspace{5ex}
\noindent {\large\bf 3.~~SL(4) Ward identities}
\vspace{3ex}

  A convenient basis for the generators in the coset space SL(4)/O(4)
consists of six matrices
$(B^{ij})^\m_{~\n} = \d^{i\m} \d^j_{~\n} + \d^{j\m} \d^i_{~\n}$,
$1\le i<j \le 4$, together with three diagonal matrices
$(B^{k})^\m_{~\n} = \d^{k\m} \d^k_{~\n} - \d^{k+1,\m}\, \d^{k+1}_\n$,
$1\le k \le 3$.
For simplicity, we will assume below that $B$ is either $B^{12}$ or
$B^1$. Notice that, considered as polarization tensors, these $B$-s lie in the
$(x^1,x^2)$-plane, and they describe transverse spin two excitations if the
only non-zero momentum components are $p_3$ and $p_4$.

  The standard derivation of a Ward identity begins with the promotion of the
constant parameter of the global transformation to a local one. We thus
consider a {\it local} transformation of the form
\beq
  y^\m_i \to (y_i')^\m = A(y_i)^\m_{~\n}\, y^\n_i\,,
  \quad\quad
  A(y) = e^{\a(y)B}\,.
\label{lcl}
\eeq
We demand that this transformation be one-to-one, \ie eq.~(\ref{lcl}) defines
a {\it diffeomorphism}.

  The requirement that $y'(y)$ be one-to-one places a non-trivial restriction
on $\a(y)$.  Unlike in any other case, we cannot assume that $\a(y)$ is a
completely arbitrary function\footnote{
As a result of this restriction, it is in general not possible
to freely interchange an insertion of $y^\m_i$ in the Ward identity with
$\partial/\partial p_\m$ is its Fourier transform.
}.
Let us consider two cases
explicitly.

\vspace{1ex}

\noindent {\it Case 1}. We assume that $\a(x) = \a_0\, \th_{\cd'}(x)$, where
$\th_{\cd'}(x) = 1$ if $x\in \cd' \subset \cd_0$, and $\th_{\cd'}(x) = 0$
otherwise. (This form should be viewed as a limiting case
of a suitable family of diffeomorphisms). As mentioned above, we let
$B=B^{12}$ or $B=B^{1}$. The mapping $y \to y'(y)$ will be one-to-one
{\it iff} $\a(x)$ is a function of $x^3$ and $x^4$ only.
For definiteness let us consider the slab $\cd'(L')$ consisting of all
points which satisfy $|x^3|,|x^4|\le L'$  for some $0<L'<L$.
We will take $\a(x) = \a_0\, \bar\th(x)$, where $\bar\th(x)=1$ if
$x\in\cd'(L')$ and $\bar\th(x)=0$ otherwise.

 The restriction to $x^3$- and $x^4$-dependence only, means that we can derive
the Ward identity only for {\it transversal} polarization. A similar situation
will be found below in the case of plane waves. Recall that under suitable
dynamical conditions, the Ward identity will imply that the correlation length
in the corresponding channel is infinite. It is encouraging that
the Ward identity is applicable in the physically relevant case of a spin two
channel.  We comment, however, that the restriction to transversal
polarization may be a technical one. It is plausible that the correlation
length is infinite also in other channels of $g^{\m\n}(x)$ that correspond
to the rest of the SL(4)/O(4) generators. The correlation length may ultimately
be infinite in all channels of $g^{\m\n}(x)$ except for its trace.

  For finite $L$, the transformation~(\ref{lcl}) with
$\a(x) =\a_0\, \bar\th(x)$ does not leave the boundaries in the $x^1$- and
$x^2$-directions invariant.
This should in principle give rise to an extra surface
term in the Ward identity. As it turns out, that extra term vanishes
identically, because $\O(y_1,\ldots)$ is zero whenever one of the internal
vertices touches the boundary.

\vspace{1ex}

\noindent {\it Case 2}. We assume that $\a(y) = \a_0\,\sc(py)$. To check when
the transformation~(\ref{lcl}) is one-to-one, let us first calculate the
jacobian $J = |\det\partial y'/\partial y|$ for general $\a(y)$. Using
\beq
  {\partial (y')^\m \over \partial y^\n} =
  A^\m_{~\n} + \partial_\n\a\, A^\m_{~\r}\, B^\r_{~\l}\, y^\l \,,
\eeq
we find
\beq
  J = \left| 1 + \partial_\m\a\, B^\m_{~\n}\, y^\n \right| \,,
\eeq
where we have used that for any two vectors $u_i$ and $v_j$,
$\det (\d_{ij} + u_i v_j) = 1 + u_j v_j$.

 Superficially, one can always force $y'(y)$ to be one-to-one simply by
taking the parameter $\a_0$ to be sufficiently small. But this is true
only for finite $L$. In general, the limit $\a_0\to 0$ involved in the
derivation of the Ward identity and the limit $L\to\infty$ will not be
interchangeable. Again, this difficulty is avoided in the case of spin two
excitations. Requiring transversality, \ie $p_\m B^\m_{~\n}=0$,
implies $J=1$ identically.

  To summarize, in both cases the transformation~(\ref{lcl}) is
a {\it volume preserving} diffeomorphism (or a limiting case of such
diffeomorphisms) which leaves the flat measure $d^4y$ invariant.

  Notice that the above transformations act on $\Tilde{Z}(\ct)$, and in this
sense the Ward identity is derived triangulation by triangulation.
One can ask why it is necessary to sum over an ensemble of triangulations
in order to obtain a theory of gravity. While the following argument
is certainly not rigorous, we believe that it captures the essential physics.

  In order to see what might go wrong if we have a
{\it fixed triangulation} model, imagine that we take one
four-simplex and we blow it up. Namely, we move  the vertices of that
four-simplex to new positions, such that the length of every link in the new
four-simplex is now huge compared to the average link length. To accommodate
this change, we will have to empty from other vertices the entire volume of the
new four-simplex. This amounts to a major deformation of the triangulation
because, before that transformation, the volume now occupied by the single huge
four-simplex typically contained many vertices.  This means that the dynamics
of a fixed-triangulation model should be non-local, {\it if the notion of
locality is defined with respect to the} $R^{(4)}$ {\it norm}.

  We comment that a ``fixed triangulation model'' can have a perfectly
consistent continuum limit which, however, describes a relativistic field
theory in {\it flat space}. One can take a periodic simplicial lattice as the
fixed triangulation, and define a matter action on its sites. In this
example, the dynamics of the matter fields is completely decoupled from the
(non-local) dynamics of the $y_i$-s. The embedding $R^{(4)}$ plays no role in
the continuum limit, and a consistent set of observables can be
defined only in terms of the lattice sites. In the random triangulation model,
on the other hand, operators that depend explicitly on the coordinates
of $R^{(4)}$, like $g^{\m\n}(x)$, should play an important role in the
continuum limit.

  In the random triangulation model, if we take
any finite number of vertices $y_k$ and we move them to arbitrary new
positions $y'_k$, then the most likely change will be relinking in the vicinity
of the old and the new positions of the vertices. The likelihood of a big
change in any region which is far away from both the $y_k$-s and the $y'_k$-s
should be negligible. Therefore, summing over an ensemble of triangulations can
give rise to dynamics which is local with respect to the $R^{(4)}$ norm.

  We are now ready to derive the SL(4) Ward identity for transversal
polarization. Let us denote by $\D_0$ the four-simplex containing
the origin as an internal point. (Recall that cases where the origin is not
an internal point of one four-simplex can be neglected).
We begin by considering the following expectation value
\beq
  G = Z^{-1} \sum_\ct e^{-S(\ct)} \prod_{i=1}^{N_0} \int_{-L/2}^{L/2} d^4y_i\,
                   \O(y_1,\ldots,y_{N_0})\, \tr B g(0)\,.
\label{G}
\eeq
Taking $\a(y)$ to be as in one of the above two cases,
the Ward identity is derived by making the change of variables~(\ref{lcl})
in eq.~(\ref{G}) and using the fact that $dG/d\a_0 = 0$ identically.
Using the invariance of the flat measure under volume preserving
diffeomorphisms, the Ward identity takes the form
\beq
  \svev{ \d_{\a(y)}\, \tr B g(0) + \tr B g(0)\, \d_{\a(y)}\log\O } = 0 \,.
\label{WI}
\eeq
Consider the first term. A straightforward
substitution gives rise to
\beq
  \svev{ \d_{\a(y)}\, g^{\m\n}(0) } = \svev{ \sum_{<ij>\in\D_0}
  \left( \d_{\a(y_i)}\, y^\m_i - \d_{\a(y_j)}\, y^\m_j \right)
  ( y^\n_i - y^\n_j ) }
  + (\m \leftrightarrow \n)  \,,
\label{dg}
\eeq
\beq
   \d_{\a(y_i)}\,y^\m_i = \a(y_i) B^\m_{~\l}\, y^\l_i \,.
\label{dy}
\eeq
Notice that we cannot take $\a(y_i)$ outside the expectation value in
eq.~(\ref{dg}), because $y_i$ is an integration variable. In other words,
the function $\a(x)$ is promoted to an operator $\a(y_i)$.

  Without being more specific, we will not be able to get any
useful information out of the Ward identity. We therefore turn now to examine
each of the above two cases separately. As we will see, one can obtain
interesting physical results, provided a certain probability distribution that
characterizes the dominant link length obeys suitable bounds.

\vspace{1ex}

\noindent {\it Case 1}. In this case we have (denoting $\d_{\a(y)}\to\d_1$)
\beq
  \d_1 \tr B g(0) = \tr B^2 (2q + r) \,,
\label{c1g}
\eeq
where
\bqry
  q^{\m\n} & = & {\sum}' \, (y^\m_i - y^\m_j) (y^\n_i - y^\n_j) \,, \\
  r^{\m\n} & = & {\sum}'' \, y^\m_i (y^\n_i - y^\n_j) +
  (\m\leftrightarrow\n)\,.
\eqry
Here ${\sum}'$ is a sum over those links of $\D_0$ that belong to the interior
of the slab $\cd'(L')$. ${\sum}''$ is a sum over the links that intersect
the boundaries of this slab, where the inner vertex is $y_i$ and
the outer vertex is $y_j$.
Notice that the $q$-term in eq.~(\ref{c1g}) is positive definite, whereas the
$r$-term is not.

  We now begin to see what are the dynamical conditions needed to obtain
interesting physical conclusions from the Ward identity. What we need is
that the dominant link length will define a stable microscopic scale.
The $r$-term should then be negligible for $L'$ which corresponds to some
macroscopic scale.

  In the forgoing discussion, it will be convenient to consider the
probability distribution $\Tilde\cp_x(l)$ for the length of the {\it longest
link} of the four-simplex containing a given point $x$.
Interesting continuum behaviour can arise only if the conditions stated below
apply uniformly, and so we will usually omit the explicit reference to $x$.
These conditions are as follows.

\noindent (a) The dynamical scale defined as
$\tilde{a}^2 = \svev{l^2}$, should satisfy $\tilde{a}/L\to 0$ in the infinite
volume limit. Here the average is taken with respect to the probability
distribution $\Tilde\cp(l)$.

\noindent (b) In the limit $l/\tilde{a}\to\infty$,  $\Tilde\cp(l)$ should
decrease faster than $(l/\tilde{a})^{-\b}$ for some $\b>3$.

  The definition of $\Tilde\cp(l)$ allows us to focus on the relevant dynamical
properties, and the scale $\tilde{a}$ characterizes the
dominant link length. The behaviour of $\Tilde\cp(l)$ depends not only
on the distribution of link lengths, but also on the volume distribution of
four-simplexes, because there is a higher probability of finding the point $x$
inside a four-simplex with a large volume. One can think of the quantity
$\G(l)=-\log\Tilde\cp(l)$ as an effective potential for the dominant link
length. An interesting question is the relation between the ``kinematical''
microscopic scale $\bar{a}$ and the dynamical scale $\tilde{a}$.
While a natural guess is $\bar{a}\approx\tilde{a}$, more complicated behaviour
could arise because of the effect of the action $S(\ct)$.

  If the above conditions are satisfied, we find for $L'\gg\tilde{a}$
\beq
   \vev{ \d_1 \tr B g(0) } = C \,,
\label{lhs}
\eeq
where
\beq
   C \equiv \svev{ \sum_{<ij>\in\D_0} l^2_{ij} } \,.
\label{vev}
\eeq
Here $l^2_{ij}=(y_i-y_j)^2$. It is clear from the above discussion that
$C\approx\tilde{a}^2$. In deriving this result we have used the cubic
symmetries and the fact that $B^2$ is a projection operator on the
$(x^1,x^2)$-plane.

  Turning now to the second term in eq.~(\ref{WI}), we find
\beq
  \d_1 \log\O =  \sum_{(i_0,\ldots,i_4)}
      \tr W^{-1}(y_{i_0},\ldots,y_{i_4})\, \d_1 W(y_{i_0},\ldots,y_{i_4})\,.
\label{dW}
\eeq
An explicit expression for $\d_1 W$ is easily obtained using eqs.~(\ref{W})
and~(\ref{dy}). One finds that $\tr W^{-1}\, \d_1 W$ vanishes, except for
those four-simplexes that intersect the boundary of the slab $\cd'(L')$.

  Putting the two terms together we obtain the Ward identity
\beq
  -\svev{ {\sum}''\, \tr W^{-1}\, \d_1 W\,\, \tr B g(0) } = C \,.
\label{WI1}
\eeq
As discussed above, eq.~(\ref{WI1}) is valid for $L'\gg\tilde{a}$, provided
there exists $\b>3$ such that $\Tilde\cp(l)$ is bounded by
$(l/\tilde{a})^{-\b}$ for $l\gg\tilde{a}$.

  The Ward identity~(\ref{WI1}) takes the form of a surface sum over a two
four-simplex correlator, where one four-simplex contains the origin, and the
other intersects the boundary of the slab $\cd'(L')$. Clearly, exponential
damping of all correlation functions beyond some finite correlation length is
incompatible with eq.~(\ref{WI1}).  Consequently, if the above mild conditions
on $\Tilde\cp(l)$ hold, the Ward identity~(\ref{WI1}) implies that {\it the
correlation length in the transversal channel of} $g^{\m\n}(x)$ {\it is
infinite}.

\vspace{1ex}

\noindent {\it Case 2}. We next examine the content of the Ward
identity in the case of plane waves. (We now let $\a(y) = \a_0\exp(ipy)$
and denote the corresponding local variation by $\d_2$).
Expanding $\exp(ipy)$ into a Taylor series and using eqs.~(\ref{dg})
and~(\ref{dy}) we find
\beq
   \vev{ \d_2 \tr B g(0) } = C + {\it Rem.}
\label{lhs2}
\eeq
The constant $C$, which represents the leading order contribution, is the
same as in eq.~(\ref{lhs}). To estimate the remainder, we use the naive bound
\beq
   \left| {\it Rem.} \right| \le \sum_{n=1}^\infty
   {p^n \svev{l^{n+2}} \over n!} \,.
\label{rem}
\eeq
Here $l$ stands for the length of the longest link of $\D_0$.

  A sufficient condition for neglecting the remainder in the
limit $p\tilde{a}\to 0$, is $\Tilde\cp(l)\sim \exp(-(l/\tilde{a})^\g)$
for some $\g>1$. This condition can be rephrased by saying that the effective
potential $\G(l)$ should rise faster than linearly.
In the case $\g<1$ inequality~(\ref{rem}) is useless, whereas the case
$\g=1$ is marginal, and convergence of the sum on the \rhs of
inequality~(\ref{rem}) depends on the subleading behaviour.

  In the case $\g>1$, or $\g=1$ plus an appropriate subleading behaviour,
one can also replace $\sin(p_\m(y^\m_i-y^\m_j))$ by $p_\m(y^\m_i-y^\m_j)$
in the second term $\d_2 \log\O$.
This give rise to the following Ward identity
\beq
  p_\m G^\m(p) = C \,, \quad\quad (p\tilde{a})^2\to 0 \,,
\label{WI2}
\eeq
where
\beq
  G^\m(p) = -i\svev{ \sum_{(i_0,\ldots,i_4)} e^{ip y_0}
            J^\m(y_{i_0},\ldots,y_{i_4})\, \tr B g(0) }\,,
\label{gp}
\eeq
and
\beq
  J^\m = \sum_{j=1}^4 \,
     (W^{-1})^j_{~\l} B^\l_{~\n}\, W^\n_{~j}\, W^\m_{~j} \,.
\label{J}
\eeq
In deriving this result we have used the transversality
condition $p_\m B^\m_{~\n}=0$ and the identity $\tr W^{-1} B W = 0$.

\vspace{5ex}
\noindent {\large\bf 4.~~Discussion}
\vspace{3ex}

  The SL(4) Ward identities~(\ref{WI1}) and~(\ref{WI2})--(\ref{J}), together
with the conditions for their validity, are the main result of this paper. The
SL(4) Ward identities constitute the link between the dynamical properties of
the discretized model and the desired continuum limit. What we need is that a
stable microscopic scale, defined by the dominant link length, will exist in
the infinite volume limit. If this condition is satisfied, we expect to obtain
a translation invariant phase where SL(4) is broken down to O(4). The unique
ground state of this phase should be flat space, and the existence of massless
spin two excitations can be regarded as a consequence of the Goldstone theorem.

  In this phase, the $R^{(4)}$ norm is a relevant concept of distance for low
energy observables. A-priori, the only concept of distance in the DT approach
is the microscopic geodesic distance, defined as the minimal number of links
between two vertices.  But if we succeed in building a phase with the above
properties, then the $R^{(4)}$ norm should regain much of the physical
significance that it has in ordinary field theories. An important property of
a consistent continuum limit is that the low energy dynamics will be
local with respect to the usual $R^{(4)}$ distance. As
long as we stay sufficiently far away from matter concentrations, the $R^{(4)}$
distance should also provide a good first approximation to the {\it macroscopic
geodesic distance} defined in terms of the expectation value of $g^{\m\n}(x)$
on some physical state. And the $R^{(4)}$ distance should coincide with the
macroscopic geodesic distance on the ground state.

  The dynamical scenario proposed in this paper is different from the one
contemplated in previous works~\cite{am,aj}. We do not attempt to identify the
continuum limit with a continuous phase transitions. Rather, the existence of
spin two Goldstone bosons should characterize an {\it entire phase} of the
model, and so every point inside that phase may be used to define the continuum
limit\footnote{
  Recently it has been suggested that, in the DT approach, the continuum limit
may be associated with regions inside the branched polymer phase~\cite{bs}.
}.

  If gravitons are Goldstone bosons, then it should be possible to derive the
associated low energy theorems, and these low energy theorems should provide an
opportunity to compare the predictions of the model with those of General
Relativity.  However, as we already explained in the introduction, carrying out
this program involves very delicate issues. After analytic continuation to
Minkowski space, the long distance limit of the correlation functions of the
model should correspond to some gauge fixed version of General Relativity. In
particular, we expect that the SL(4) currents will not be generally covariant
in that limit. As a result, uncovering the physical content of the low energy
theorems, requires one to investigate what are the physical observables of
the model. Understanding the properties of the {\it energy-momentum}
tensor may play an important role in the construction of a
consistent set of observables.

  Apart from the  Maison-Reeh theorem~\cite{mr}, there is another argument
which indicates that the SL(4)/O(4) broken phase can only correspond to a
gauge-fixed version of General Relativity.  The diffeomorphism group is {\it
locally non-compact}, and the local value of the metric tensor $g^{\m\n}(x)$ is
an element of the coset space GL(4)/O(4). Even if we suppress the conformal
degree of freedom, the corresponding coset space SL(4)/O(4), now considered as
the space of values of $g^{\m\n}(x)$ with a fixed trace, is still non-compact.
As a result, obtaining a manifestly diffeomorphism invariant version of General
Relativity in the long distance limit, is incompatible with the finiteness of
the expectation value of $g^{\m\n}(x)$. We comment that this behaviour might
correspond in some formal sense to a ``topological phase'' characterized by
uncontrolable stretching of simplexes.

  An alternative line of investigation, is to look for a direct relation
between the long distance limit of the euclidean correlation functions of the
model and some gauge fixed version of {\it euclidean} gravity. To date,
however, there is no widely agreed solution to the question of how the
conformal factor should be treated in continuum euclidean
gravity~\cite{ghp,mm}.
This means that there is no generally agreed answer to the question of what is
a consistent gauge fixed version of euclidean gravity. The dynamics of the
present model singles out the trace of the metric tensor from the rest
of its component. Thus, we hope that further investigations of the model may
shed new light on this difficult issue.

  One can also try to construct a manifestly unitary model, at the price of
sacrificing manifest O(4) invariance~\cite{y}. To this end, one should first
divide the hypercube $\cd_0\subset R^{(4)}$ into equal time slices. The
rules for constructing triangulations on each time slice should be analogous
to the ones used in this paper, and an extra rule has to be provided for the
linking between neighbouring time slices. The global symmetries of this
formulation should include discrete-time and continuous-space translations,
as well as SL(3) transformations. The existence of physical spin two Goldstone
bosons can then be attributed to the spontaneous breakdown of SL(3) to O(3).
In this formulation one should prove the recovery of full Lorentz covariance
in the low energy limit.

\vspace{5ex}
\noindent {\large\bf 5.~~Dynamical considerations and the choice of the action}
\vspace{3ex}

  At this stage, we do not know whether the condition of
faster-than-exponential damping of $\Tilde\cp(l)$ can be satisfied for some
range of parameters. (We also leave open the possibility that a
weaker condition may be sufficient for the existence of momentum eigenstates).
Numerical simulations can clearly help us in getting some idea about the
structure of the phase diagram, as well as about the behaviour of
$\Tilde\cp(l)$ in different phases.
Here we only intend to give a very preliminary description of
the issues in question, and to point out to possible relations between the
choice of $S(\ct)$, the abstract triangulation's action, and the desired
dynamics.

  In principle, there may be two potential sources for an unsuppressed
behaviour of $\Tilde\cp(l)$ at large $l$. One possibility is that the
shape of four-simplexes might fluctuate very strongly on small scales.
Here the term ``small scale'' refers to the microscopic geodesic distance.
In other words, we are asking what is the probability of finding a very
elongated four-simplex (whose volume is $O(1)$) only a few links away from
regular-looking four-simplexes.

  Experimenting with geometrical triangulations suggests that big
``short distance'' fluctuations in the shape of simplexes should be strongly
damped. We first observe that a very elongated four-simplex can never appear
alone in a region of regular-shape four-simplexes. There must
exist a transition region containing more and more elongated four-simplexes.
If the length of the longest link in that region is $l_0 \gg 1$,
then the typical number of four-simplexes in the transition region should
be $O(l_0)$ or larger.

  Moreover, consider two four-simplexes sharing a common tetrahedron, and
assume that one of them is much more stretched than the other. Namely, assume
that the longest link of one of them is longer than the longest link of the
other by a large factor $z\gg 1$. Under these circumstances, the phase space
available for the common vertices will typically be reduced by a factor of
$O(1/z)$. The origin of this reduction is the need to  keep the volume of {\it
both} four-simplexes $O(1)$. If we assume even a finite suppression factor for
every pair of four-simplexes in the transition region, we arrive at an overall
exponential suppression factor as a function of $l_0$.  Needless to say, it
will take a more serious investigation to determine what is the actual
behaviour of $\Tilde\cp(l)$. But one can at least hope that the condition of
faster-than-exponential damping is not totally unreasonable.

  There is another potential source for an undamped behaviour of
$\Tilde\cp(l)$, which is related to the long range dynamics. Let us assume that
strong short distance fluctuations in the length of links are indeed
suppressed.  Thus, when we start near the boundary, we should find mainly
regular-shape simplexes, and the quasi-local average of every entry of
$g^{\m\n}$ should be $O(1)$.

  However, the local SL(4) order parameter may gradually change as we move
away from the boundary. What is the probability that we will ultimately
reach regions where {\it all} four-simplexes are extremely stretched in
one direction and extremely squeezed in another direction? If this happens,
then we are facing an infra-red instability, and $\Tilde\cp(l)$ will
show no damping at all at large $l$. In this case we can expect no
more than the trivial constraint $\Tilde\cp(l)=0$ for $l>2L$.

  The answer to the above question may depend critically on the dimensionality
of the model. What we have described above is nothing but the familiar
mechanism for the restoration of continuous symmetries in {\it two dimensions}.
It is plausible that the same physical mechanism should work in
a two dimensional version of the present model as well.

  How can this be reconciled with our previous argument that the expectation
value of $g^{\m\n}$ necessarily break SL(d) down to O(d)? The resolution of
this paradox is that a {\it well-defined} SL(d)-{\it symmetric phase does not
exist}. The reason is that, as mentioned earlier, the coset space SL(d)/O(d) is
non-compact. Consequently, a presumed SL(d)-symmetric phase would have to be
dominated by infinitely stretched simplexes.

  A two dimensional version of the present model may be unstable against long
range variations in the local SL(2) order parameter, as in the case of any
other continuous symmetry in two dimensions.  If this is indeed the case, then,
away from the boundary, all triangulations will be dominated by extremely
stretched triangles, whose length will be limited only by the finite size of
the embedding square.  As the infinite volume limit is taken, one may obtain a
phase where most of the triangles have a finite distance from the boundary.
That ``topological'' phase will be neither translationally invariant nor
SL(2)-symmetric\footnote{
  If this behaviour is verified, it would imply that a two
dimensional version of the present model is no exception to the ``$c=1$
barrier''~\cite{a}.
}.

   The infra-red instability described above is known to be a peculiar property
of two dimensional kinematics. There is no obvious reason to suspect that our
{\it four dimensional} model will suffer from a similar instability.
But, going to four dimensions does not automatically guarantee the absence
of an infra-red instability. The latter could occur if for some reason the
would-be Goldstone modes developed a $1/p^4$ behaviour instead of the
usual $1/p^2$ pole.  The issue will have to be investigated in the future,
alongside with the effect of short distance fluctuations on $\Tilde\cp(l)$.

  How do these considerations affect the choice of the action $S(\ct)$?
In our model, the metric tensor $g^{\m\n}(x)$ is formally a composite field,
and there is no obvious advantage in using the Regge discretization of the
Einstein-Hilbert action. (By analogy, we do not expect to find an explicit
kinetic term for pions in the QCD lagrangian). Since we will not use the Regge
action, we also avoid the unboundedness problem that may aflict that action in
the infinite volume limit.

  A possible choice of the action is
\beq
   S(\ct) = \k_4 N_4 + \g' \sum_i \left( \cn_4(i) - \bar\cn_4 \right)^2\,.
\label{S}
\eeq
Here $\cn_4(i)$ denotes the number of four-simplexes sharing the
$i$-th vertex, and $\bar\cn_4$ is some average value of $\cn_4(i)$.
As usual, $N_4$ is the total number of four-simplexes. The first term on the
\rhs of eq.~(\ref{S}) is the familiar cosmological term, whereas the second
term suppresses the occurrence of high $\cn_4$ vertices. The action~(\ref{S})
is positive definite for $\k_4,\g'\ge 0$. Notice that the action~(\ref{S}) is
local. On the other hand, the term $(N_4-V)^2$ which is commonly added to the
Regge action in numerical simulations is not, because one cannot write
$(N_4-V)^2$ as a single sum over a local quantity.

  The grand-canonical partition function is well-defined only if the number of
triangulations is exponentially bounded as a function of $N_4$.  At present, it
is unclear whether or not the exponential bound exists in the ensemble of
abstract four-triangulations with a fixed topology~\cite{ckr,aj2,bs}.
The restriction to geometrical triangulations may carry the extra benefit of
allowing us to establish an exponential bound. The infinite volume limit should
then be defined by letting $\k_4\to \k^c_4$ from above. (Otherwise, one would
have to use the canonical partition function. In that case it is convenient to
take $N_0$ as the independent variable, and to define the infinite volume limit
by letting $N_0\to\infty$).

  We propose to include the second term in the action for the following reason.
We believe that there is a correlation between strong short distance
fluctuations in the SL(4) order parameter, and the occurrence of  high $\cn_4$
vertices.  The tendency for the buildup of a high $\cn_4$ increases as we try
to decrease the size of the transition region between an extremely long
four-simplex and a surrounding of regular four-simplexes. The increased
likelihood for high $\cn_4$ vertices in the transition region, was taken into
account in our previous argument concerning the expected suppression of short
distance fluctuations in the SL(4) order parameter.  But if the resulting
large-$l$ damping of $\Tilde\cp(l)$ turns out to be marginal, then the extra
necessary damping may be provided by the second term in eq.~(\ref{S}).

   A related issue is the role of {\it crumpled} triangulations in the present
model\footnote{
There is some evidence \cite{bs,aj2} that, in the ensemble of abstract
triangulations with a four-sphere topology, crumpled triangulation
overwhelm in the limit $N_4\to\infty$.
}.
Crumpled triangulations are characterized by the existence of vertices with
very high coordination number (or, equivalently, very high $\cn_4$). The high
connectivity may prevent many abstract crumpled triangulations from having a
realization in $R^{(4)}$. In particular, the occurrence of vertices whose
$\cn_4$ is of the same order of magnitude as the total number of four-simplexes
should be suppressed.  Moreover, if an abstract crumpled triangulation does
have a realization in $R^{(4)}$, then every high $\cn_4$ vertex, as well as
every vertex which is a neighbour of a high $\cn_4$ vertex, should typically
have a very limited room to move around. We thus expect that the contribution
of crumpled triangulations to the partition function will be suppressed in the
present model from the outset.

  Apart from the parameters that enter the action eq.~(\ref{S}), the
model contains an additional free parameter. This parameter is the power of
$s$  in eq.~(\ref{F}).  The results of this paper remain valid in the more
general case $F(s)=\th(s)\,s^n$ for $n>1$, with the minor change of a factor
of $n$ in the appropriate places (\eg the \rhs of eq.~(\ref{dW})).
The role of the parameter $n$ may resemble the role of the inverse temperature
in classical statistical mechanics. If true, then this parameter may have
a significant effect on the dynamics of the model.

  In this paper we made no attempt to carry out a systematic investigation of
the phase diagram of the model. We focused on the physical properties that
should characterize a translation invariant phase where SL(4) is spontaneously
broken to O(4). Another phase having a stable microscopic scale may be
characterized by further breaking of rotation and translation invariance to a
discrete subgroup. This would be the case if the ground state resembles a
regular lattice. Although we expect a different (and richer) massless spectrum
is that phase, it may also give rise to a physically interesting continuum
limit. Another possibility is that of a ``topological phase'' characterized by
a
divergent link length in the infinite volume limit. Future investigations
should
help us in getting a more detailed understanding of the model and its phase
diagram, and in deciding whether the dynamical scenario proposed in this paper
is viable.

\vspace{5ex}
\noindent{\bf Acknowledgment}
\vspace{3ex}

  I thank A.\ Casher, S.\ Elitzur, T.\ Jacobson and A.\ Schwimmer for
discussions and for useful comments.

%%%%%%%%%%%%%%%%%%%%%%%%%%%%%%%%%%%%%%%%%%%

\vspace{5ex}
\centerline{\rule{5cm}{.3mm}}

\end{document}